\documentclass[twocolumn,pra,amsmath,amssymb,showkeys,showpacs]{revtex4}% Physical Review A

\usepackage{graphicx}% Include figure files
\usepackage{dcolumn}% Align table columns on decimal point
\usepackage{bm}% bold math

\newcommand*{\del}{\delta}
\newcommand*{\thet}{\theta}
\newcommand*{\mO}{\text{O}}
\newcommand*{\smO}{\text{{\tiny O}}}

\begin{document}

\title{Hyperspherical elliptic coordinates treatment of muon transfer from muonic hydrogen to atomic oxygen} 
\author{Arnaud Dupays}
\affiliation{
Laboratoire Collisions, Agrégats, Réactivité, IRSAMC,
Universit\'e P. Sabatier, 31062 Toulouse, France
}
\author{Bruno Lepetit}
\affiliation{
Laboratoire Collisions, Agrégats, Réactivité, IRSAMC,
Universit\'e P. Sabatier, 31062 Toulouse, France
}
\author{J.\ Alberto\ Beswick}
%\email{beswick@irsamc.ups-tlse.fr}
%\homepage[Visit: ]{www.car8.ups-tlse.fr}
\affiliation{
Laboratoire Collisions, Agrégats, Réactivité, IRSAMC,
Universit\'e P. Sabatier, 31062 Toulouse, France
}
\author{Carlo Rizzo}
\affiliation{
Laboratoire Collisions, Agrégats, Réactivité, IRSAMC,
Universit\'e P. Sabatier, 31062 Toulouse, France
}%
\author{
Dimitar Bakalov}
\affiliation{%
INRNE, Bulgarian Academy of Sciences, Sofia, Bulgaria
}%
%\date{\today}

%
\begin{abstract}
Quantum-mechanical calculations of muon transfer between muonic
hydrogen and an oxygen nuclei 
for $s$ waves and
collision  energies in the range $10^{-3} - 10^3$ eV, are presented.
Close-coupling time-independent Schr\"odinger equations,
written in terms of hyperspherical elliptic coordinates were integrated
along the hyper-radius to obtain the partial and total muon-transfer
probabilities. 
The results show the expected Wigner-Bethe threshold behavior up
to  collision energies of the order of $10^{-2}$ eV and 
pronounced maxima at $10^2$ eV which can be interpreted in
terms of  crossings between potential energy curves corresponding
to the  entrance channel
state $(\mu p)_{1s} + \mO$ and two product channels which asymptotically
correlate to $p + (\mO\mu)_{n=5,6}$.
 The population of the final states with different orbital angular momenta
is found to be essentially independent of energy in the range considered in this work. This can be attributed 
to a strong  selection rule for the conservation of the quantum
number associated to one of the elliptic hyperangles.
\end{abstract}

\pacs{36.10.Dr}
%
%\keywords{muonic atoms}
\maketitle

\section{\label{Intro}Introduction} 

Negative muon transfer between muonic atoms (muonic hydrogen, for instance) 
and other atoms or molecules has been extensively studied in the framework
of muon catalyzed nuclear fusion (see Ref.~\cite{Pono:01} and literature cited therein). Also, the structural and 
spectroscopic
properties of these species
are of interest for metrology and in
 tests of quantum electrodynamics \cite{Baka:93,Dupa:03b}. 

Recently, several theoretical  \cite{Baka:93,Sult:00,Adam:01,Sult:02,Dupa:03a} 
and experimental \cite{Wert:98} works have considered the
problem of muon transfer from the muonic hydrogen to an oxygen molecule. 
Since
the muonic hydrogen has to approach one of the oxygen  nuclei
very close in order for the muon to be transferred  \cite{Gers:63}, 
 the process can be described as
\begin{equation}\label{I1}
(p\mu)_{1s}+\mO^{8+}\to p+(\mu\mO)^{7+}_{n\ell}.
\end{equation}
 
Although there have been several full three-dimensional calculations
of muon transfer rates at low energies between muonic-hydrogen and low-$Z$ atoms
(see literature cited in Ref.~\cite{Sult:02}),
there is none when the transfer involves nuclei with $Z>3$ . 
Indeed, as  $Z$
 increases  there is a larger initial-channel polarization
and a stronger final-channel Coulomb interaction
which make the full quantum calculation computationally
heavy.  Thus up to now only approximate calculations have
been performed for the muon-transfer rate between muonic-hydrogen and 
oxygen \cite{Haff:77,Gers:63,Sult:00,Dupa:03a}. 
We present here the first  numerically converged three-dimensional calculations
for reaction (\ref{I1}) for $s$ waves and collision
energies in the range $10^{-3} - 10^{3}$ eV.  Since for  the entrance channel 
the centrifugal barrier
for $J=1$ is about $0.1$ eV, the calculations presented here can be considered
as full 3-dimensional 
up to thermal energies. 

The calculation were performed as follows.
Hyperspherical elliptic coordinates \cite{Tols:95,Tols:01}
have been used.
A piecewise diabatic basis set on the hyperspherical angles
 was used to expand the wave function.
The resulting close-coupling time-independent
Schr\"odinger equations in the hyper-radius were solved using a de Vogelaere algorithm
and the partial and total muon-transfer probabilities were determined by the standard
S-matrix analysis at large distances.
Since for energies below $10^{-1}$ eV, the muon-transfer process studied here is equivalent
to an ultra-cold collision (de Broglie wavelength, $\lambda>1$ \AA, much larger
than the effective range, $a\sim 0.1$ \AA, of the potential interaction), special care had to be
taken to the asymptotic analysis in the entrance channel.

The paper is organized as follows. Section II introduces the model and the methodology
used in the calculations. Section III presents the calculated muon-transfer probabilities
together with 
their  interpretation in terms
of simple Landau-Zener and threshold models. 
Finally, section IV is devoted to the conclusions.

%%%%%
%%%%%

\section{Methodology}

\begin{figure}
\includegraphics[width=8.cm]{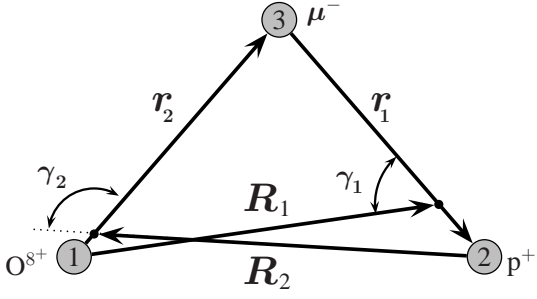}   
\caption{\label{figure_a2_1}{The two sets of mass-scaled Jacobi coordinates
corresponding to the entrance and exit channel of reaction (\ref{I1}).
The center of mass positions are not in scale.}}
\end{figure}

We start with the two mass-scaled Jacobi sets of coordinates ($\bm R_1, \bm r_1$)
and ($\bm R_2, \bm r_2$) adapted to
the entrance and exit channels of reaction (\ref{I1}),
respectively (see fig. \ref{figure_a2_1}).
They are defined by 
\begin{eqnarray}\label{M1a}
\bm R_i &=& \sqrt{\frac{m_{i,jk}}{m}}\,\left(\frac{m_j\,\bm x_j + m_k\,\bm x_k}{m_j+m_k} -\bm x_i 
\right),
\\ \label{M1b}
\bm r_{i} &=& \sqrt{\frac{m_{j,k}}{m}}\,(\bm x_j - \bm x_k),
\end{eqnarray}
where the set $(i=1,j=2,k=3)$ corresponds to (O, p, $\mu$), and $(i=2, j=3, k=1)$ to
(p, $\mu$, O). The reduced masses $m_{i,jk}$ and $m_{j,k}$ in Eqs.\ (\ref{M1a}) and (\ref{M1b}), are given by
\begin{equation}\label{M2a}
m_{i,jk} = \frac{m_i\,(m_j + m_k)}{m_{i} + m_j + m_k},
\quad m_{j,k} =  \frac{m_j\,m_k}{ m_j + m_k},
\end{equation}
while $m$ is choosen as
\begin{equation}\label{M2b}
m = \left(
\frac{m_{i}\,m_j\,m_k}{m_{i} + m_j + m_k}
\right)^{1/2}.
\end{equation}
These sets are related by the orthogonal transformation
\begin{equation}\label{M3}
\left( \begin{array}{c}\bm R_2\\ \bm r_2\end{array} \right)
= \left( \begin{array}{cc}
-\cos\theta_\mu & -\sin\theta_\mu \\ \sin\theta_\mu & -\cos\theta_\mu
\end{array} \right)\,\left( \begin{array}{c}\bm R_1\\ \bm r_1\end{array} \right),
\end{equation}
where 
\begin{equation}\label{M4}
\tan\theta_\mu = \frac{m_\mu}{m},
\end{equation}
giving in our case $ \theta_\mu\simeq 19^\circ$. 

In spherical coordinates, the system can be described by three
Euler angles specifying the overall orientation, the two distances
($R_i, r_i$) and the angle $\gamma_i$ between the two vectors
$\bm R_i$ and $\bm r_i$ (see figure \ref{figure_a2_1}).
%The volume element is the given by
%\begin{equation}\label{v1}
%dV = \rho^5\,d\rho\,\pi^2\,\sin^2(\chi_i)\,\sin(\theta_i)\,d\chi_i\,d\theta_i
%\end{equation}
Two sets of Delves hyperspherical coordinates are then defined by
the common hyper-radius
\begin{equation}\label{M5}
\rho = \sqrt{R_{i}^2 + r_{i}^2}
\end{equation}
the hyper-angles
\begin{equation}\label{M6}
\tan(\chi_{i}/2) = \frac{r_{i}}{R_{i}};\quad i=1,2.
\end{equation}
and the $\gamma_i$ angles.

The relationship between the two sets of Delves shape angles is given by
%\begin{equation}\label{M7}
\begin{eqnarray}
&&\left( \begin{array}{c} 
\cos\chi_2\\ \sin\chi_2\,\cos\gamma_2\\ \sin\chi_2\,\sin\gamma_2
\end{array}
 \right)
\nonumber\\
&= &\left( \begin{array}{ccc} 
\cos(2\theta_\mu) & \sin(2\theta_\mu)&0 \\ -\sin(2\theta_\mu) & \cos(2\theta_\mu)& 0\\0&0&1
\end{array} \right)\left( \begin{array}{c}
\cos\chi_1\\ \sin\chi_1\,\cos\gamma_1\\ \sin\chi_1\,\sin\gamma_1
\end{array} \right).
%\end{equation}
\label{M7}
\end{eqnarray}

The hyperspherical elliptic coordinates are defined by
\begin{subequations}
\label{M8}
\begin{eqnarray}\label{M8a}
\eta &=& \chi_1 - \chi_2,\quad -2\,\theta_\mu\leq\eta\leq2\,\theta_\mu,
\\ \label{M8b}
\xi &=& \chi_1 + \chi_2,\quad 2\,\theta_\mu\leq\xi\leq 2\,\pi - 2\,\theta_\mu,
\end{eqnarray}
\end{subequations}
with the volume element
\begin{equation}\label{v2}
d\tau = \rho^5\,d\rho\,\frac{\pi^2}{4\sin(2\theta_\mu)}\,(\cos(\eta)-\cos(\xi))\,d\eta\,d\xi.
\end{equation}

In terms of these
coordinates the kinetic energy operator for total angular momentum zero, is given by
\begin{widetext}
%\begin{eqnarray}
%T &=&
%-\frac{\hbar^2}{2m}\,\Bigg(\frac{1}{\rho^5}\,
%\frac{\partial}{\partial\rho}\,\rho^5\,\frac{\partial}{\partial\rho} \nonumber\\
%&+& \frac{16}{\rho^2}\,\frac{1}{\cos(\eta)-\cos(\xi)}\,
%\bigg[\frac{\partial}{\partial\eta}\Big(\cos(\eta)-\cos(2\theta_\mu)\Big)\,\frac{\partial}{\partial\eta}
% \\&-&
%\frac{\partial}{\partial\xi}\Big(\cos(\xi)-\cos(2\theta_\mu)\Big)\,\frac{\partial}{\partial\xi}\bigg]
%\Bigg)\nonumber
%\label{M9} 
%\end{eqnarray}
\begin{equation}
T=-\frac{\hbar^2}{2m}\,\Bigg(\frac{1}{\rho^5}\,
\frac{\partial}{\partial\rho}\,\rho^5\,\frac{\partial}{\partial\rho} 
+ \frac{16}{\rho^2}\,\frac{1}{\cos(\eta)-\cos(\xi)}\,
\bigg[\frac{\partial}{\partial\eta}\bigg(\cos(\eta)-\cos(2\theta_\mu)\bigg)\,\frac{\partial}{\partial\eta}
-
\frac{\partial}{\partial\xi}\bigg(\cos(\xi)-\cos(2\theta_\mu)\bigg)\,\frac{\partial}{\partial\xi}\bigg]
\Bigg),
\label{M9}
\end{equation} 
%and the
% coulombic interaction potential is given by $V(\rho,\eta,\xi) = e^2\,C(\eta,\xi)/\rho$, with
%\begin{eqnarray}
%C(\eta,\xi) = 
%
%-\frac{e^2\,Z_{\smO}}{\vert \bm x_\mu - \bm x_{\smO}\vert}
%+ 
%\frac{e^2\,Z_{\smO}}{\vert \bm x_p - \bm x_{\smO}\vert} 
%- \frac{e^2}{\vert \bm x_p - \bm x_\mu\vert},
%
%-\frac{m^{1/2}_{\mu,\smO}\,Z_{\smO}}{s\lbrack(\xi+\eta)/4\rbrack}
%-\frac{m^{1/2}_{\mu,p}}{s\lbrack(\xi-\eta)/4\rbrack}
%4\,\frac{c(\eta/2)+c(\xi/2)}{\cos(\eta)-\cos(\xi)}\,
%\bigg[z^+\,c(\eta/4)\,s(\xi/4) +
%z^-\,s(\eta/4)\,c(\xi/4)\bigg]
%+\frac{(2m_{\smO,p})^{1/2}\,Z_{\smO}}{
%\sqrt{1 + p^+\,c(\eta/2)\,c(\xi/2)-p^-\,s(\eta/2)\,s(\xi/2)}}
%+Z_{\smO}\,\sqrt{\frac{2m_{\smO,p}}{
%1 + p^+\,c(\eta/2)\,c(\xi/2)-p^-\,s(\eta/2)\,s(\xi/2)}}
%\label{M10}
%\end{eqnarray}
\end{widetext}
%where
%\begin{equation}\label{M11}
%p^+ = 1 + \frac{2\,m_\mu}{m_p+m_{\smO}},\quad p^-=\frac{m_p-m_{\smO}}{m_p+m_{\smO}},
%\end{equation}
%and 
%\begin{equation}\label{M12}
%z^{\pm} = Z_{\smO}\,m^{1/2}_{\mu,\smO}\pm m^{1/2}_{\mu,p}.
%\end{equation}

For a given value of the hyper-radius $\rho$, the total wave function $\psi(\rho,\eta,\xi)$ is expanded in terms of 
a basis set of $N_{ch}$ functions 
$\phi_i(\eta,\xi;\rho)$ depending on the hyperspherical angles
$\eta$ and $\xi$.
We use a diabatic-by-sector
representation. In each sector $\rho_n-\del\rho_n \leq \rho < \rho_n+\del\rho_n; n=1,...,N_{\rho}$
we write: 

\begin{equation}\label{M13}
\psi(\rho,\eta,\xi) = \frac{1}{\rho^{5/2}}\,\sum_{i=1}^{N_{ch}}\,F_i(\rho)\,\phi_i(\eta,\xi;\rho_n) 
\end{equation}
where $\phi_i(\eta,\xi;\rho_n)$ are eigenstates of the Hamiltonian at fixed $\rho_n$ distances.
Their calculation requires the solution of a bound state problem for the Coulomb  potential
 
\begin{equation}\label{M14}
V=-\frac{e^2}{\vert \bm x_p -\bm x_\mu\vert} - \frac{8\,e^2}{\vert \bm x_\mu -\bm x_{\smO}\vert}
+ \frac{8\,e^2}{\vert \bm x_p -\bm x_{\smO}\vert}
\end{equation}
 presenting two attractive 
singularities at $(\eta,\xi)=(\pm2\theta_\mu,2\theta_\mu)$ corresponding  to a vanishing
muon-oxygen and muon-proton distance. 
%Treatment of Coulomb singularities requires 
%the use of specific methods such as increasing the density of grid points, or alternatively the 
%oscillation frequency of the basis functions, 
%near the singularities. Mapped Fourier \cite{Fatt:96, Lemo:00}, 
%Lagrange \cite{Baye:99} or Schwartz \cite{Schw:85,Duns:02}
% interpolations are examples of such methods which have been used for
%cases with one singularity. 

% In terms of the hyperspherical elliptic coordinates the two attractive terms
%in the Coulomb potential $V$
%can be written in terms of a prefactor $(\cos(\eta)-\cos(\xi))^{-1}$. 
%Therefore,
%since the kinetic energy operator $T$ has also this factor in the
%denominator we can write the eigenproblem

This bound state problem can be rewritten as:
\begin{equation}\label{M15}
\left\{-\frac{16\hbar^2}{2m\rho^2_n}\,[\hat L(\eta) - \hat L(\xi)]
+ W(\rho_n,\eta,\xi)\right\}\,\phi_i(\eta,\xi;\rho_n)=0,
\end{equation}
where
\begin{equation}\label{M16}
\hat L(u) = \frac{\partial}{\partial u}\big(\cos u-\cos(2\theta_\mu)\big)\,\frac{\partial}{\partial u}
\end{equation}
and
\begin{equation}\label{M17}
 W(\rho_n,\eta,\xi) = [\cos(\eta)-\cos(\xi)]\,[V(\rho_n,\eta,\xi)-\epsilon_i(\rho_n)].
\end{equation}

Equation (\ref{M15}) can be viewed as a zero eigenvalue problem depending
parametrically on the potential $[\cos(\eta)-\cos(\xi)]\,[V(\rho_n,\eta,\xi)-\epsilon_i(\rho_n)]$, 
in the sense that the constant $\epsilon_i(\rho_n)$ is being adjust in such a way that the operator
on the left hand side of equation (\ref{M15}) has zero eigenvalue. 
The renormalized potential $W(\rho_n,\eta,\xi)$ has two important features. One is that
the two Coulomb singularities are regularized, the other being that it is
approximately separable. We can therefore write

\begin{equation}
W(\rho_n,\eta,\xi)= 
W_\eta(\rho_n,\eta)+W_\xi(\rho_n,\xi) + \Delta W(\rho_n,\eta,\xi),
\label{M17b}
\end{equation}
where $W_\eta(\rho_n,\eta) = W(\rho_n,\eta,2\,\theta_\mu)$ and
$W_\xi(\rho_n,\xi) = W(\rho_n,-2\,\theta_\mu,\xi) - W(\rho_n,-2\,\theta_\mu,2\,\theta_\mu)$.

\begin{figure}
\includegraphics[width=8.cm]{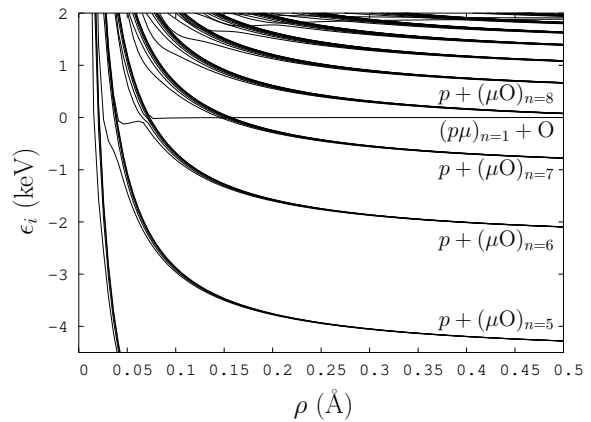}
\caption{\label{figure_a2_2}{Calculated $\epsilon_i$ energies of the
hyperspherical elliptic basis as a function of the
hyper-radius.}}
\end{figure}

The 2-dimensional 
problem can be solved by exploiting this approximate separability of the potential. 
For instance, defining $\eta=2\,\theta_\mu\,\bar\eta$
with $-1\leq\bar\eta\leq 1$, we get from equation (\ref{M15})
\begin{widetext}
\begin{equation}\label{M18}
\Bigg[-\frac{4\,\hbar^2}{m\,\rho_n^2} \,\frac{s^2(\theta_\mu)}{\theta^2_\mu}\,
\frac{\partial}{\partial\,\bar\eta}\,\bigg(1 -
\frac{\sin^2\theta_\mu\,\bar\eta}{\sin^2\theta_\mu}
\bigg)\,\frac{\partial}{\partial\,\bar\eta}
+ W_\eta(\rho_n,\eta)\Bigg]\,\varphi_k(\eta;\rho_n) =\epsilon^{(\eta)}_{k}(\rho_n)\,\varphi_k(\eta;\rho_n)
\end{equation}
\end{widetext}
The similarity between the differential operator in (\ref{M18}) (in particular in the limit $\thet_\mu\to0$) 
with the one defining Legendre polynomials, suggests to use the latter as basis set functions
for expansion of $\varphi_k(\eta;\rho_n)$.
We solve a similar problem for $\xi$ using the $W_\xi(\rho_n,\xi)$ potential and
obtain $\varphi_\ell(\xi;\rho_n)$ eigenfunctions and the $\epsilon^{(\xi)}_{\ell}(\rho_n)$
eigenvalues. We then iterate (using a bi-section method on $\epsilon_i$) until
we get $\epsilon^{(\eta)}_{k}(\rho_n)+\epsilon^{(\xi)}_{\ell}(\rho_n)=0$. Once the separable basis
set is obtained, solutions of (\ref{M15}) for the full non-separable potential are obtained
by diagonalizing the representation matrix of the full Hamiltonian in the product basis.
In Fig.~\ref{figure_a2_2} we present the calculated energies $\epsilon_i$  as a function of the
hyper-radius $\rho$.  The origin of energies has been chosen
to coincide with the asymptotic limit of the entrance channel
$(p\,\mu)_{n=1} + \mO$. The calculations presented in this work cover the
energy range between this limit and the $p + (\mu\,\mO)_{n=10}$
threshold at about 1 keV. 

The $N_{ch}$ coupled equations are integrated
along the hyper-radius $\rho$ using the de Vogelaere algorithm \cite{Lepe:86}. 
This provides a logarithmic derivative matrix $Z$ at $\rho_M=\rho_{N_\rho} + \delta_{N_\rho}$.
For the energy range considered here, we included 88 channels:
($(p\,\mu)_{n=1-2} + \mO$ and
$p + (\mu\,\mO)_{n=1-9}$).  The integration of the coupled equations
was performed from the origin to
 $\rho_{\mbox{end}}\sim 200\,a_\mu\sim 1$ \AA.
The asymptotic analysis has been performed using
the appropriate Jacobi coordinates 
for the entrance  and   for the product channels.
Elementary asymptotic wave functions for the different final arrangement channels $\lambda=1,2$  are written 
as products of translational functions $f_{n\ell}(R_\lambda)$, Coulomb bound wave functions
$C_{n\ell}(r_\lambda)$ and normalized Legendre polynomials:
\begin{equation}\label{M19}
\psi_{n\ell\lambda}(R_\lambda,r_\lambda,\gamma_\lambda) = f_{n\ell}(R_\lambda)\,C_{n\ell}(r_\lambda)\,
\bar P_\ell(\cos\gamma_\lambda).
\end{equation}
We use these elementary asymptotic wave functions to form $N_{ch}$ physical solutions
whose  forms are given by
\begin{eqnarray}
\Psi^{(a)} &=& \psi^S_{n\ell\lambda}(R_\lambda,r_\lambda,\gamma_\lambda)
\nonumber\\
&+&\sum_{n'\ell'\lambda'}\,\psi^C_{n'\ell'\lambda'}(R_{\lambda'},r_{\lambda'},\gamma_{\lambda'})\,
K_{n'\ell'\lambda'\leftarrow n\ell\lambda}
\label{M20}
\end{eqnarray}
where the superscripts $S$ and $C$ refer to the type of translational functions ($S$ for sine-type
functions for open channels and exponentiallly decaying functions for closed channels, $C$
for cosine-type and exponentially growing functions, respectively). If a Coulombic interaction
is still important (as for instance in the product channels O$\mu + p$ in our problem), Coulomb wave functions
are used instead. 

We define the matrices $\bm F$ and $\bm F'$ as the projections of the elementary asymptotic functions
on the $\phi_i$ hyperspherical basis. This projection is performed at the maximum
hyperspherical radius $\rho_M$
\begin{widetext}
\begin{subequations}\label{M21}
\begin{eqnarray}
F_{i,n\ell\lambda} &=&  \langle \phi_i \vert \psi_{n\ell\lambda}\rangle_{\rho=\rho_M}
%\nonumber\\
= \int\,d\eta\,d\xi\,\left[\cos(\eta)-\cos(\xi)\right]\,\phi_i(\eta,\xi;\rho_M)\,\psi_{n\ell\lambda}(R_\lambda,
r_\lambda,\gamma_\lambda)\vert_{\rho_M}
\label{M21a} \\
F'_{i,n\ell\lambda} &=& \langle \phi_i \vert \partial\psi_{n\ell\lambda}/\partial \rho\rangle_{\rho=\rho_M}
%\nonumber\\
=\int\,d\eta\,d\xi\,\left[\cos(\eta)-\cos(\xi)\right]\,\phi_i(\eta,\xi;\rho_M)\,\partial\psi_{n\ell\lambda}(R_\lambda,
r_\lambda,\gamma_\lambda)/\partial\rho\vert_{\rho_M}
\label{M21b}
\end{eqnarray}
\end{subequations}
\end{widetext}
The $N_{ch}$ linearly independent solutions which result from the propagation
steps up to $\rho_M$ are linear combinations of the $N_{ch}$ asymptotic solutions
given by equations (\ref{M21}). This can be restated as an equality of the
logarithmic derivative matrix $\bm Z$ in the hyperspherical basis
\begin{equation}\label{M22}
\bm Z = ({\bm F'}^S + {\bm F'}^C\,\bm K) (\bm F^S + \bm F^C\,\bm K)^{-1}
\end{equation}
The $\bm K$ matrix, and then the $\bm S$ matrix, can be extracted from (\ref{M22}).
%%%%
 \section{Results of the calculations}
  
We have performed calculations 
of the reaction
\begin{equation}\label{R1}
(p\mu)_{n=1}+\mO^{8+}\to p+(\mu\mO)^{7+}_{n,\ell}
\end{equation}
for collision energies in the range $10^{-3}-10^{3}$ eV and for
total angular momentum $J=0$ ($s$ waves).
For large $R$ distance between the oxygen nuclei and the proton, the potential for the entrance channel
of reaction (\ref{R1})
behaves as $V=-\alpha\,Z^2e^2/2R^4$, where $Z=8$ and $\alpha = (9/2)(\hbar^2/m_{\mu,p}e^2)^3$.
Since with this potential the centrifugal barrier has 
a height of $(\hbar^2J(J+1)/m_{\smO,p\mu}Ze)^2/8\alpha$,  for energies below $0.1$ eV 
the partial wave $J=0$ is the only one which contributes to the
cross sections. Thus for thermal energies these calculations are essentially exact full 3-dimensional.

\begin{figure}
\includegraphics[width=8.cm]{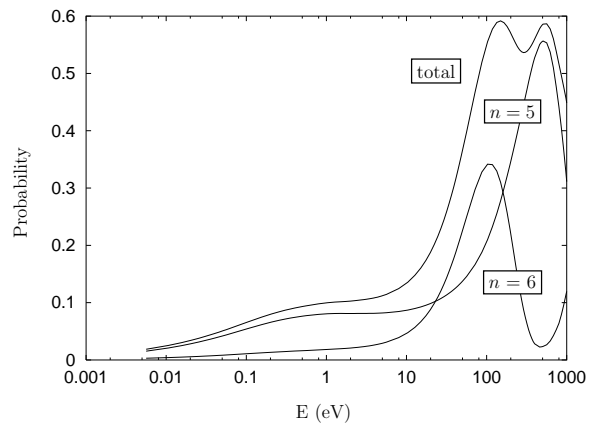}
\caption{\label{figure_a2_3}{Partial and total muon-transfer probabilities for 
reaction (\ref{R1}) as a function of the
collision energy.}}
\end{figure}

Fig.\ \ref{figure_a2_3} presents  the total probabilities for muon transfer
as a function of the energy, as well as the partial muon-transfer probabilities into the
 $p + (\mu\,\mO)_{n=6}$ and $p + (\mu\,\mO)_{n=5}$ channels. The other open channels have negligeable
populations. This is in complete agreement with what was found in
approximate calculations \cite{Gers:63}. 
These results  can be qualitatively understood by inspection
of figure \ref{figure_a2_2}. Starting in channel $(p\,\mu)_{n=1} + \mO$,
the system crosses diabatically the channel 
 $p + (\mu\,\mO)_{n=7}$. Muon transfer is completely negligible 
as the coupling is very small compared with the collision energy.
The couplings  to  channels $p + (\mu\,\mO)_{n=6}$ and $p + (\mu\,\mO)_{n=5}$ are 
larger as evidenced by  avoided
crossings.  The other
channels $p + (\mu\,\mO)_{n<5}$ are weakly coupled to the initial one
and they are not expected to be populated significantly.  

\begin{figure}
\includegraphics[width=8.cm]{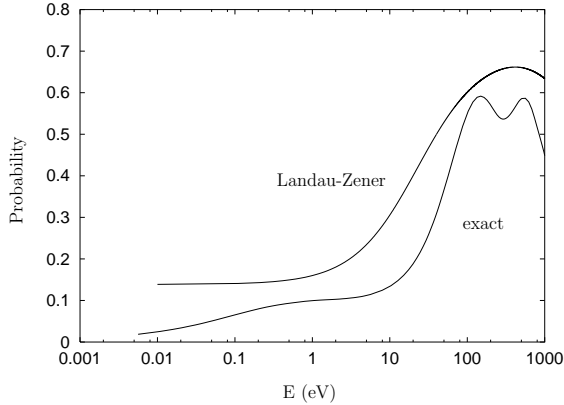}
\caption{\label{figure_a2_3b}{Total muon-transfer probabilities for 
reaction (\ref{R1}) as a function of the
collision energy using a simple three channel problem and Landau-Zener
semi-classical 
approximation.}}
\end{figure}

Actually, a simple calculation involving only three channels ($(p\,\mu)_{n=1} + \mO$,
$p + (\mu\,\mO)_{n=5-6,\ell=0}$)  and using the semi-classical Landau-Zener formula for the 
two crossings provides  qualitative agreement with the exact results down
to energies of the order of 1 eV (see figure \ref{figure_a2_3b}). For lower energies,
quantum threshold behavior dominates the semi-classical Landau-Zener approximation
and for very low energies Wigner's threshold laws predict a $E^{1/2}$
dependence.
This low energy region corresponds to $k\,a \ll1$, with $k = (2\,m_{\smO,p\mu}\,E)^{1/2}/\hbar$
and $a$ being the range of the potential in the entrance channel.
Defining the effective range of the potential by $V(a)\leq E$, we get $E\ll 0.1$ eV. This is exactly
what it is found in our calculations (see figure \ref{figure_a2_3}).

\begin{figure}
\includegraphics[width=8.cm]{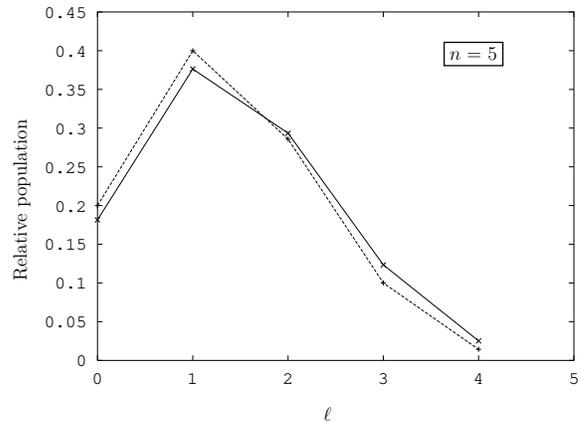}   
\caption{\label{figure_a2_4}{Exact (full line) and approximate (dotted) relative populations of the $\ell$ levels for $n=5$ and collision energy
$E = 0.1$ eV.}}
\end{figure}

\begin{figure}
\includegraphics[width=8.cm]{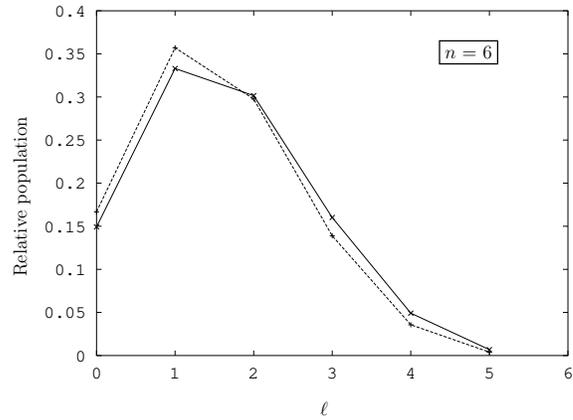}   
\caption{\label{figure_a2_5}{Exact (full line) and approximate (dotted) relative populations of the $\ell$ levels for $n=6$ and collision energy
$E = 0.1$ eV.}}
\end{figure}

In figures \ref{figure_a2_4} and \ref{figure_a2_5} we present relative populations
of O$\mu_{n=5-6,\ell}$ levels for collision energy
$E=0.1$ eV. This distribution is almost independent of collision energy
in the energy range considered here. This fact and the particular form
of the distributions in figures \ref{figure_a2_4} and \ref{figure_a2_5} can be
understood as follows. From an inspection of figure \ref{figure_a2_2} it is clear that
only one of the set of states correlating asymptotically to a given O$\mu_n$ manifold is significantly coupled
to the initial $p\mu_{n=1}$ state. Due to the approximate separability of the
potential, each of these states can be assigned to approximate hyperspherical
elliptic quantum numbers $(n_\eta,n_\xi)$. Since the initial $p\mu_{n=1}$ channel is
colinearly dominated and $n_\xi>0$ corresponds to excitation away from the
colinear configuration, the final states significantly coupled to the initial channel
satisfy the approximate selection rule: $n_\xi=0$. At large distances, hyperspherical
coordinates tend gradually toward parabolic coordinates\cite{Tols:01}. The final $(n_\eta,
n_\xi=0)$ states can therefore be identified to $(\bar n_\eta,\bar n_\xi=0)$ states
with $\bar n_\eta = n-1$ and $\bar n_\xi =n_\xi)$. 
%Since the final kinetic energy
%is large, these states undergo a sudden
%transformation into the asymptotic free atoms states.
The final approximate $\ell$ distribution is
then given by the square of the projection of this states onto the final $(n,\ell)$ spherical
basis set with the result \cite{Park:60}:
\begin{equation}
\label{Res1}
P^{app}_{n\ell} \propto \vert C ((n-1)/2,(n-1)/2,-n/2,-n/2;\ell,0)\vert^2
\end{equation}
where $C$ stands for a Clebsh-Gordan coefficient.
We have represented this approximate distribution in figures \ref{figure_a2_4} and \ref{figure_a2_5}.
Clearly, the $n_\xi=0$ selection rule is a very good approximation.

\begin{figure}
\includegraphics[width=8.cm]{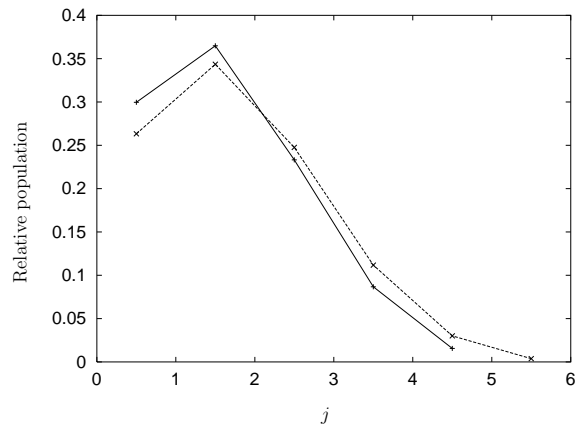}   
\caption{\label{figure_a2_6}{Relative populations of the $j$ levels for $n=5$ (dotted)
and $n=6$ (full) for a collision energy
$E = 0.1$ eV.}}
\end{figure}

Experimentally only the populations of the fine-structure levels $j$ are eventually
measurable. Thus, from the orbital angular momentum $\ell$ probabilities  one gets
\begin{equation}
P_{n,j} \propto \sum_\ell\, \frac{P_{n,\ell}}{2\ell+1}\,
\sum_{m_j,m_\ell}\,\vert C (\ell,1/2,m_\ell,m_j-m_\ell;j,m_j)\vert^2
\end{equation}
The results are presented in figure \ref{figure_a2_6}.

\section{Conclusions}

We have presented 3-dimensional calculations of muon transfer probabilities 
between muonic hydrogen and oxygen for relative translational  energies between
$10^{-3}$ and $10^3$ eV and total angular
momentum $J=0$ ($s$ waves). Since the centrifugal barrier 
for $J=1$ is of the order of 0.1 eV, for thermal and  lower energies the
calculations are essentially exact.

The results for the total and partial probabilities can be interpreted qualitatively
by a simple three channels model and Landau-Zener semi-classical approximation for
energies between 1 and 1000 eV. For energies below 0.1 eV, the probabilities follow
the expected Wigner's threshold laws behavior. 

The fine-structure levels populations are essentially independent of energy
and almost independent of the principal quantum number $n$. This 
has been attributed to an approximate selection rule for the elliptic hyperspherical
quantum number $n_\xi$.

In conclusion, the hyperspherical elliptic coordinates 
introduced by Tolstikhin \emph{et al} \cite{Tols:95}, are 
particularly well suited to the treatment of this type of reaction.

\begin{acknowledgments}

We thank  IDRIS (CNRS computing center)
for an allocation of CPU time on a NEC SX5 vector processor, and NATO for a collaborative linkage grant 
PST.CLG.978454 between Bulgaria and France.  

\end{acknowledgments}

\end{document}